\documentclass[review]{elsarticle}

\usepackage{hyperref,siunitx}
\usepackage{color,soul}
\usepackage{graphicx,subfig}

\graphicspath{{figures/png/}}

\def\PM{\ensuremath{\pm}}

\begin{document}

\journal{Magnetic Resonance Imaging}
	
\begin{frontmatter}
	
\title{Evaluation of Diffusion Weighted Imaging in the Context of Multi-Parametric MRI of the Prostate in the assessment of suspected low volume prostatic carcinoma}

\author[a,b]{Ioannis Papadopoulos \corref{d}}
\author[b]{Jonathan Phillips}
\author[b]{Rhodri Evans}
\author[c]{Neil Fenn}
\author[a]{Sophie Shermer}
\address[a]{College of Science (Physics), Swansea University, Singleton Park, Swansea SA2 8PP, United Kingdom}
\address[b]{Institute of Life Science, Medical School, Swansea University, Singleton Park, Swansea SA2 8PP, United Kingdom}
\address[c]{Morriston hospital, Heol Maes Eglwys, Morriston, Swansea SA6 6NL}
\cortext[d]{Corresponding author. \\ \textit{Email address}: ipapadopoulo@gmail.com}

\begin{abstract}	
	Data from a multi-parametric MRI study of patients with possible early-stage prostate cancer was assessed with a view to creating an efficient clinical protocol.  Based on a correlation analysis suggesting that diffusion-weighted imaging (DWI) scores are more strongly correlated with overall PIRADS scores than other modalities such as dynamic contrast enhanced imaging or spectroscopy, we investigate the combination of T2-weighted imaging (T2w) and DWI as a potential diagnostic tool for prostate cancer detection, staging and guided biopsies. Quantification of the noise floor in the DWI images and careful fitting of the data  suggests that the mono-exponential model provides a very good fit to the data and there is no evidence of non-Gaussian diffusion for $b$-values up to \SI{1000}{s/mm^2}. This precludes the use of kurtosis or other non-Gaussian measures as a biomarker for prostate cancer in our case. However, the ADC scores for healthy and probably malignant regions are significantly lower for the latter in all 20 but one patient. The results suggest that a simplified mp-MRI protocol combining T2w and DWI may be a good compromise for a cost and time efficient, early-stage prostate cancer diagnostic programme, combining robust MR biomarkers for prostate cancer that can be reliably quantified and appear well-suited for general clinical practice.
\end{abstract}


\end{frontmatter}


\section{Introduction}

Prostate cancer is the second most common cancer in men worldwide \cite{ferlay2015cancer}. For most developed countries there has been a decrease in death rates mainly due to earlier diagnosis combined with better staging and treatment \cite{center2012international, siegel2015cancer}. The diagnostic pathway has been driven by the use of serum prostate specific antigen (PSA) testing with subsequent transrectal ultrasound (TRUS) scan and prostate biopsies. However, the positive predictive value of a raised PSA and subsequent TRUS guided biopsies is low \cite{tanimoto2007prostate}, which leads to a considerable number of men undergoing unnecessary biopsies and a marked increase in the detection of clinically insignificant cancer \cite{loch2004transrectal}. This has significant cost implications and exposes patients to the inherent risk of biopsy and treatment \cite{caverly2016presentation}.

Diffusion-weighted imaging (DWI) is a functional magnetic resonance imaging (MRI) technique that takes advantage of the diffusion of water molecules in tissue to obtain information about tissue microstructure by calculating quantities such as apparent diffusion coefficient (ADC), which have been shown to be inversely correlated with increased cellularity in different tumor types \cite{perez2016diffusion}.  DWI has shown potential in differentiating malignant from benign prostate tissue, but it is usually used in combination with $T_2$-weighted imaging (T2w) and other MRI techniques for diagnostic purposes. Tissues with restricted diffusion tend to also have lower signal on T2w images \cite{Simpkin2013}. The combination with T2w and advanced MRI techniques such as DWI, dynamic contrast enhanced imaging (DCE) and spectroscopy (MRS), known as multi-parametric MRI (mp-MRI), can improve diagnosis and staging for the peripheral prostate zone \cite{delongchamps2011multiparametric}.

The aim of mp-MRI is three-fold: Potentially avoid prostate biopsies when these are unnecessary; Target biopsies particularly in men who have previously undergone a negative transrectal ultrasound scan and prostate biopsy and finally to stage the disease in line with National Institute For Health and Clinical Excellence (NICE) guidelines \cite{nice2008prostate} when patients have been diagnosed and entering an active surveillance protocol. In addition there is some evidence to suggest that these advanced MR techniques are useful in detection of low volume low grade prostate cancers that require no active treatment and are often perceived as being over diagnosis \cite{ahmed2017diagnostic}. There is also the prospect of cancer localization in patients with elevated PSA who have several negative biopsies \cite{anastasiadis2006mri, amsellem2005negative, pondman2008mr} using the greater anatomic information provided by mp-MRI.  To standardize reporting of mp-MRI findings and reduce variability, a global system for the assessment and reporting of prostate cancer (PIRADS) was proposed \cite{barentsz2012esur}, combining T2w, DWI, DCE and MRS, although MRS (alongside with other techniques) was excluded in the latest version (PIRADS v2), but may be incorporated in later versions \cite{weinreb2016pi}. Motivated by a statistical analysis of existing patient data showing strong correlation of DWI and overall PIRADS scores, we analyse the robustness of DWI-based biomarkers and their utility in improving the assessment and management of patients with suspected early-stage prostate cancer.

\section{Materials and methods}

\subsection{Patient population}

From a cohort of patients participating in a early-stage prostate cancer assessment examination, data of 46 acquired between 22/02/2015 and 24/08/2016, were selected to evaluate the robustness and utility of advanced MRI techniques. Patients were selected on three clinical criteria: (a) suitability for an active surveillance programme, (b) patients with rising PSA in whom biopsy seemed to be high risk, e.g., anti-coagulated patients and (c) patients with rising PSA who were unable to undergo biopsy for other clinical reasons such as abdominoperineal resection. The patients were between 48 and 85 years old with a mean age of 68. The study was funded by the Swansea Prostate Cancer Research Fund, ethical approval for the analysis of the data was obtained and the study was reviewed favourably by the institution's Joint Scientific and Research Committee.

\subsection{MR protocol}

In accordance with the original PIRADS (v1) protocol \cite{barentsz2012esur}, the examinations included T2w, MRS, DWI and DCE performed on a 3-Tesla MR scanner (Siemens 3T Magnetom Skyra) using a 18-channel pelvic phased-array receive coil. T2w imaging was performed in transverse, sagittal and coronal planes. T2w images were acquired using a fast spin echo (FSE) sequence with the following parameters: TR: \SI{4320}{ms}, TE: \SI{101}{\milli\second}, field of view (FOV): $256 \times 256$mm, acquisition matrix: $320 \times 320$, flip angle: $160^\circ$, slice thickness: 3mm, number of averages: 3. DWI images were acquired using transverse echo planar imaging (EPI), TR: \SI{3800}{ms}, TE: \SI{64}{ms}, FOV: $256 \times 256$mm, acquisition matrix: $128 \times 128$, flip angle: $90^\circ$, slice thickness: \SI{3}{mm}, number of averages: 8;  $b$-values: 0, 100, 800 and 1000 \si{s/mm^{2}}. 

\subsection{Data analysis}

The MR images were assessed by an experienced radiologist.  26 of the patients selected were considered to have no abnormalities or benign conditions e.g. benign prostatic hyperplasia.  For the remaining 20 patients with suspected cancer, the radiologist delineated suspicious and normal regions using custom-shaped regions of interest (ROIs) based on the T2w images.  The same ROIs were applied to the ADC data sets after conversion to adjust for differences in image resolution.  Post-processing and calculations were performed using in-house code written in \texttt{MATLAB}. 

To assess background noise levels for the signal-to-noise (SNR) analysis in the diffusion-weighted images, a ROI above the patient's abdomen was selected.  It was not possible to place a ROI for every patient due to artifacts in some of the images that made it impossible to select a suitable ROI in the air-only region above the patient's abdomen.  To evaluate the level and variability of the noise for different slices and $b$-values, the mean signal intensity was computed for all slices and $b$-values for 28 patients.

ADCs for the normal and suspicious regions were calculated by applying a least squares linear fit of the logarithm of the signal $S$ as a function of the $b$-value according to

\begin{equation}
\label{eq:ADCeq}
\ln S = \ln S_0 - b D,
\end{equation}
where $S_0$ is the signal at $b$ = 0 and $D$ is the ADC.

The same fitting routine was used to calculate ADCs for each voxel inside the prostate for all 46 patients and to generate ADC maps.   To visualize the signal decay as a function of $b$-value for all voxels inside the prostate and assess its linearity on a logarithmic scale, 3D waterfall plots were generated.  To elucidate the distribution of ADC values inside the prostate, ADC histograms were computed by placing the data in bins of  \SI{50}{mm^2/s}. The quality of each fit for both single voxel and mean ROI signal fits was assessed by calculating $R^2$, the square of the correlation between the response values $\hat{y}_i$ and the predicted response values $\bar{y}_i$. $R^2$ is a measure of how well the fit models the variation in the data.  It is defined as the ratio of the sum of squares of the regression (SSR) and the total sum of squares (SST)

\begin{equation}
  \label{eq:R2eq}
  R^2 = \frac{\mathcal{E}_{\rm SSR}}{\mathcal{E}_{\rm SST}}=1-\frac{\mathcal{E}_{\rm SSE}}{\mathcal{E}_{\rm SST}},
\end{equation}

where $\mathcal{E}_{\rm SSR}$ is defined as

\begin{equation}
\label{eq:SSR}
\mathcal{E}_{\rm SSR} = \sum_{i=1}^n w_i (\hat{y}_i -\bar{y} )^2,
\end{equation}

and $\mathcal{E}_{\rm SST}$ is the sum of squares about the mean

\begin{equation}
\label{eq:SST}
\mathcal{E}_{\rm SST} = \sum_{i=1}^{n} w_i (y_i - \bar{y})^2.
\end{equation}

$\mathcal{E}_{\rm SSE}$ is the sum of squares due to error and $\mathcal{E}_{\rm SST}$ = $\mathcal{E}_{\rm SSR}$ + $\mathcal{E}_{\rm SSE}$.   $R^2$  can take on values between $0$ and $1$.  A value closer to $1$ is indicative of a greater proportion of variance accounted for by the model.

The PIRADS scoring was performed by the same experienced radiologist by evaluating the data for each technique and assigning a value between 1 and 5 according to the level of significance. The overall PIRADS score indicates the probability of malignancy on imaging criteria. To assess the correlation of PIRADS scores for different MR techniques and overall PIRADS scores, and quantify the potential prognostic value of each technique with reference to the overall PIRADS score, Pearson correlation coefficients between $-1$ and $+1$ were computed, $+1$ indicating perfect positive, $-1$ perfect negative and $0$ no linear correlation~\cite{weinreb2016pi, pearson1895note}.   

\section{Results}

\subsection{Correlation of PIRADS Component Scores vs Total Scores}

\begin{table}

	\centering

	\begin{tabular}{|c|*{5}{c|}}\hline

        & Total  & T2      & DWI    & MRS    & DCE\\\hline
          
Total   & 1	     & 0.7633  & 0.8685 & 0.6540 & 0.7294\\\hline

T2		&        & 1	   & 0.6477 & 0.4724 & 0.3091\\\hline	

DWI		&        &         & 1      & 0.3474 & 0.5686\\\hline

MRS		&        &         &        & 1      & 0.2354\\\hline

DCE		&        &         &        &        & 1\\\hline

\end{tabular}

\caption{Correlation of T2w, DWI, MRS and DCE PIRADS scores with total PIRADS scores. DWI appears to be most reliably correlated with the overall PIRADS scores, whereas the correlation for the MRS scores is approximately 65\%.}

	\label{Table1}

\end{table}

The Pearson correlation coefficients for the component and overall PIRADS scores in Table~\ref{Table1}) show that the scores for T2w and DWI correlate more strongly with the overall PIRADS scores compared to those for MRS and DCE imaging.  Specifically, the DWI component scores and overall PIRADS scores have a Pearson score of 87\%, whereas MRS does not exceed 66\%. Interestingly, there is almost no correlation between T2w and MRS or DCE.  The lack of correlation with the overall PIRADS score, suggests that, in this cohort at least, it does not add diagnostic value.

\subsection{A posteriori background noise analysis}

\begin{table}
	\centering
	\begin{tabular}{|*{7}{c|}}

		\hline

		P 41 & P 42 & P 51 & P 55 & P 59 & P 65 & P 76 \\[-1.5ex] 

		6.0 \PM 1.3 & 4.3 \PM 1.2 & 5.6 \PM 0.6 & 4.6 \PM 0.6 & 5.3 \PM 0.5 & 6.0 \PM 0.2 & 5.8 \PM 0.3 \\\hline

		P106 & P107 & P108 & P109 & P110 & P114 & P116 \\[-1.5ex] 

		5.1 \PM 0.3 & 5.1 \PM 0.3 & 5.0 \PM 0.3 & 5.1 \PM 0.4 & 5.4 \PM 0.5 & 5.6 \PM 0.3 & 5.3 \PM 0.5 \\\hline

		P117 & P119 & P120 & P124 & P127 & P129 & P130 \\[-1.5ex] 

		5.6 \PM 0.4 & 4.3 \PM 1.0 & 4.7 \PM 0.3 & 6.0 \PM 1.2 & 5.1 \PM 0.8 & 5.2 \PM 0.4 & 5.8 \PM 0.3 \\\hline

		P131 & P132 & P133 & P134 & P136 & P137 & P139 \\[-1.5ex] 

		5.4 \PM 0.3 & 6.1 \PM 1.2 & 5.4 \PM 0.8 & 5.0 \PM 0.4  & 5.7 \PM 0.5 & 6.3 \PM 0.6 & 5.9 \PM 1.2 \\\hline

	\end{tabular}
	\caption{Noise statistics (mean/std) for selected patients}
	\label{Table2}
\end{table}

\begin{figure}
	\subfloat[]{\includegraphics[width=0.51\textwidth]{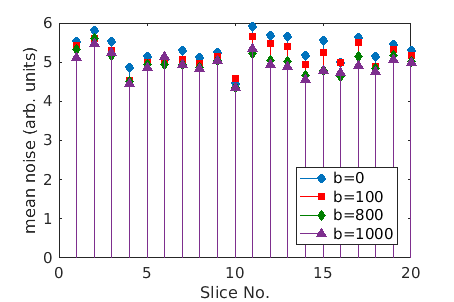}} 
     \subfloat[]{\includegraphics[width=0.51\textwidth]{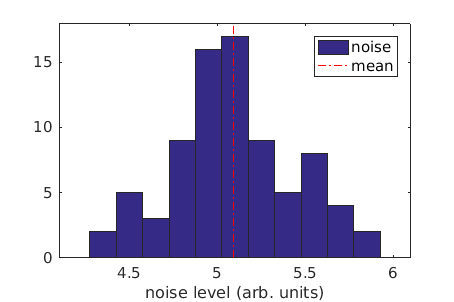}}
	\caption{(a) Noise stem plot showing mean noise level for each slice and b-value for one patient. The mean noise over ROI does not vary significantly with slice and b-value. (b) Corresponding histogram with indication of mean and standard deviation of noise.}
	\label{stem-hist113}
\end{figure}

Noise levels were fairly low with the mean signal intensity in the noise ROIs exceeding $6$ (in arbitrary units) for only a few points (Table~\ref{Table1}).   A typical noise stem plot and noise distribution histogram is shown in Fig.~\ref{stem-hist113}.  Fig.~\ref{stem-hist113} (a) shows the noise level for each slice for four different $b$-values in different colors.  We observe that the noise levels do not vary significantly with slice position and $b$-value as expected. Fig.~\ref{stem-hist113} (b) shows that the background noise level is more or less normally distributed around a mean value (red dashed line).

Based on the resulting estimates for the noise level, the threshold for significance was set to $10$, i.e., any points with signal intensity below $10$ were excluded from fits.  For the purposes of reporting by the radiologist, the ADC maps generated by the scanner software were used.

\subsection{ADC Maps, Distribution of ADC values \& Quality of Fit}

\begin{figure}
\subfloat[]{\includegraphics[scale=0.32]{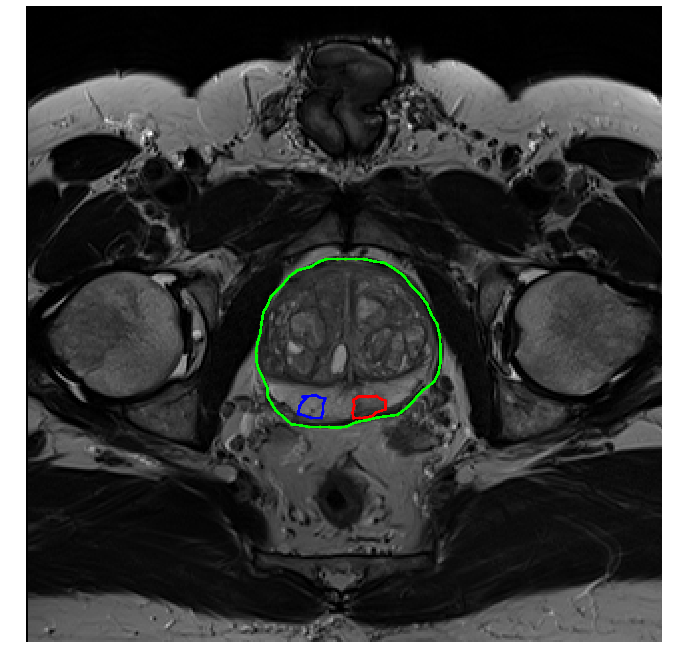}}
\subfloat[]{\includegraphics[scale=0.32]{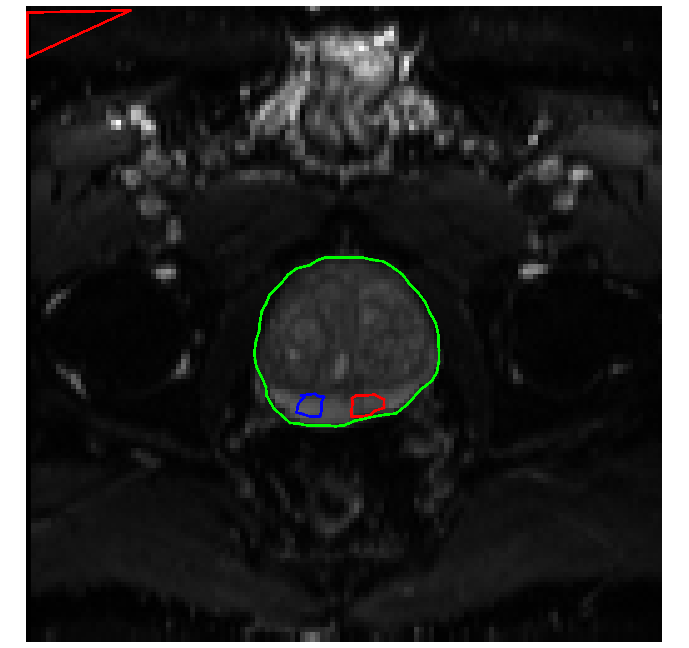}}
\caption{(a): T2w image of a selected patient within the prostate (green), the suspicious (red) and contralateral (blue) normal areas delineated. (b): These ROIs were applied to the same slice of the ADC map, in addition to a triangular ROI (top left red) for the noise analysis.}  
\label{DelineatiatedROIs113}
\end{figure}

\begin{figure}
\subfloat[]{\includegraphics[width=0.44\textwidth]{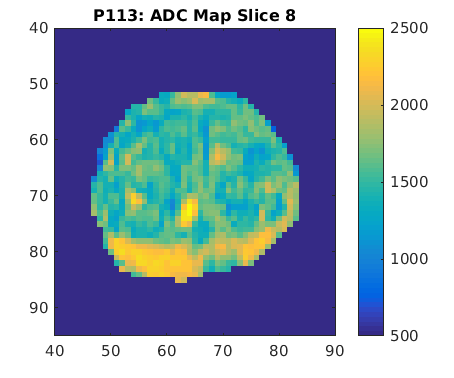}}
\subfloat[]{\includegraphics[width=0.55\textwidth]{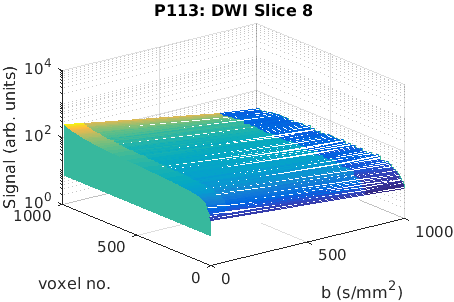}}
\caption{(a): ADC map for a delineated prostate.  (b): Signal vs $b$-values waterfall plot showing that the signal is well above the background noise level for all voxels and $b$-values and the signal decay on a logarithm scale is linear.}
\label{ADCmap113}
\end{figure}
 
\begin{figure}
\subfloat[]{\includegraphics[width=0.49\textwidth]{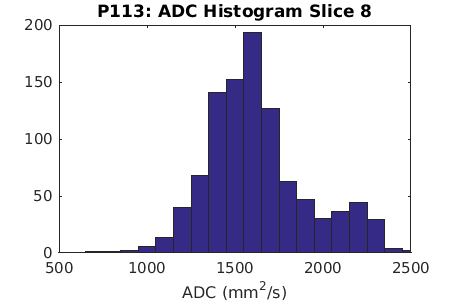}}
\subfloat[]{\includegraphics[width=0.49\textwidth]{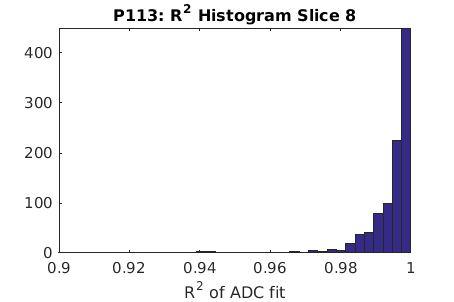}}
\caption{(a): Distribution of ADC values for one slice of a prostate. The distribution is approximately symmetric around 1577mm$^2$/s. (b): Distribution of $R^2$ values.} \label{ADChist113}
\end{figure}

ADC maps were computed for all patients and all slices.   An example of a typical ADC map is shown in Fig.~\ref{ADCmap113} (a).  A typical distribution of ADC values for voxels inside the prostate is shown in Fig.~\ref{ADChist113} (b). The respective histogram of the corresponding correlation coefficient $R^2$ (Fig.~\ref{ADChist113} (b)) shows that $R^2$ is $\ge 0.95$ for almost all voxels, suggesting that the linear monoexponential decay model provides a good fit for the data.  The linearity of the signal decay for all voxels is further evidenced by 3D waterfall plots, a typical example of which is shown in Fig.~\ref{ADCmap113} (b).  The plot clearly shows that the logarithm of a signal intensity varies linearly with $b$ for the range of values \SIrange{0}{1000}{s/mm^2} for all voxels, i.e., we observe no evidence of non-Gaussian diffusion.  This is not too surprising as usually a minimum $b$-value of \SI{1500}{s/mm^2} is required to quantify non-Gaussianity~\cite{phillips2015simple}.  In the waterfall plots  brighter colors (yellow) indicate higher signal and darker colors (blue) lower signal.

\subsection{ADC values for normal and suspicious ROI for patients}

\begin{figure}
\subfloat[]{\includegraphics[width=0.49\textwidth]{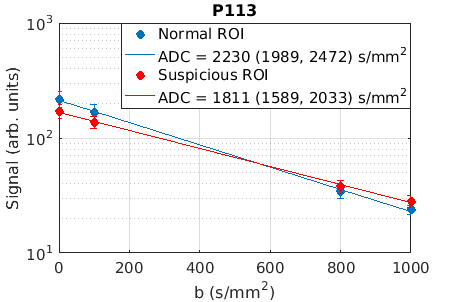}}
\subfloat[]{\includegraphics[width=0.49\textwidth]{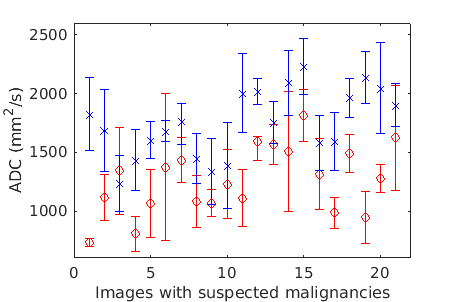}}
\caption{(a): $b$-value to signal intensity semi-logarithmic plot. The linear ADC fit is indicative of no kurtosis for both suspicious and normal delineated regions. (b): Normal (blue) Vs Suspicious (red) ADC scatter plot, confirms that normal regions have higher ADC values than suspicious ones. The error bars are relatively high for the suspicious regions which corresponds to higher variation of the diffusion in this area.} \label{bfit113}
\end{figure}

Fig.~\ref{bfit113} shows a typical signal vs $b$-value plots for a normal and a suspicious ROI for a patient with suspected cancer. The signal decays faster, i.e., the ADC is higher, in the normal region.  In both cases the linear fit is excellent, the signal remains well above the noise floor for all $b$-values and the variation of the signal over the respective ROI is small. Table~\ref{Table3} shows the ADC for the normal and suspicious regions for the 20 patients with suspected cancer, showing that the ADC for the suspicious ROIs is lower than the value for the normal ROI except for one outlier.

\begin{table}
	\centering
	\begin{tabular}{|c|*{3}{c|}}
		\hline
		 &   ADC (mm$^2$/s) &  \\\hline
		Patient&  Suspicious ROIs& Normal ROIs\\\hline
		P015  & 731 (699, 762)	 & 1821 (1511, 2132)\\\hline
		P038  & 1114 (919, 1309) & 1686 (1336, 2036)\\\hline
		P041  & 1342 (971, 1714) & 1233 (993, 1473)\\\hline
		P048  & 807 (658,  955)  & 1434 (1175, 1692) \\\hline
		P051  & 1065 (777, 1354) & 1602 (1445, 1760)\\\hline
		P059  & 1373 (745, 2000) & 1680 (1590, 1770) \\\hline
		P090  & 1434 (1245, 1623)& 1761 (1565, 1917)\\\hline
		P094  & 1081 (862, 1301) & 1447 (1236, 1658) \\\hline
		P095  & 1066 (949, 1183) & 1334 (1054, 1614) \\\hline
		P099  & 1228 (934, 1522) & 1385 (1054, 1614) \\\hline
		P105  & 1109 (866, 1353) & 2004 (1688, 2339)\\\hline
		P110a & 1592 (1427, 1632)& 2020 (1910, 2129)\\\hline
		P110b & 1566 (1394, 1738)& 1754 (1578, 1930)\\\hline
		P112  & 1508 (1000, 2016)& 2091 (1812, 2369)\\\hline
		P113  & 1811 (1589, 2033)& 2230 (1989, 2472)\\\hline
		P118  & 1313 (1010, 1616)& 1581 (1346, 1817)\\\hline
		P119  & 984 (851, 1118)  & 1594 (1346, 1842)\\\hline
		P120  & 1490 (1331, 1649)& 1962 (1797, 2126)\\\hline
		P129  & 944 (723, 1164)  & 2137 (1913, 2361)\\\hline
		P134  & 1276 (1157, 1395)& 2045 (1656, 2434)\\\hline
		P139  & 1622 (1173, 2072)& 1902 (1715, 2089)\\\hline		
	\end{tabular}
	\caption{Mean ADC with 95\% confidence intervals for 20 patients with suspected malignancies.}
	\label{Table3}
\end{table}

\section{Discussion}

The combination of advanced MR techniques (mp-MRI) has become a useful tool in the management of patients at risk of or diagnosed with prostate cancer \cite{Fuetterer2015}. It is highly important for staging and has the potential to assist therapy guidance in the future \cite{turkbey2010prostate}. According to Ahmed et al. \cite{ahmed2017diagnostic} prostate biopsies could be avoided by a large proportion of men if mp-MRI was applied for screening.  Isebaert et al. \cite{isebaert2013multiparametric} showed that the combination of T2w, DWI and DCE increased sensitivity.   Tan et al. \cite{Tan2015} reported that DCE combined with T2w improved prostate cancer detection but Hansford et al. \cite{hansford2015dynamic} demonstrated low differentiating value between cancerous and healthy regions in the prostate especially in early-stage cancers.  Our result suggest that DWI has the highest correlation with overall PIRADS scores and our preliminary data on DWI combined with T2w imaging in a cohort of patients selected for mp-MRI screening gives promising results in terms of discrimination between normal and suspicious tissue.  This raises the possibility of tailoring mp-MRI protocols for screening purposes.  Dropping DCE and MRS significantly reduces the overall examination time, in our case from 55 to 25 minutes, resulting in considerable cost savings, improved patient experience and elimination of potential short and long-term risks associated with exposure to gadolinium-based contrast agents \cite{kanda2013high}.

Except for one, 
all probably malignant cases (based on PIRADS scoring) had higher ADC values for suspicious ROIs than respective normal ROIs (Fig.\ref{Table2}), in line with previous studies \cite{chen2008prostate,hosseinzadeh2004endorectal,issa2002vivo}.  The ADC of the suspicious ROIs ranged from \SIrange{944}{1811}{mm^2/s} (mean{\PM}std 1448\PM \SI{272}{mm^2/s}) and those of the normal ROIs ranged from \SIrange{1514}{2230}{mm^2/s} (mean{\PM}std 1910\PM \SI{229}{mm^2/s}). The error bars for the suspicious regions are significantly larger than the respective normal ones, indicative of more variation in the diffusion values and ADCs in these regions. The wide range of values obtained (for patients P113 and P119 it is almost twice as much for the suspicious regions) is consistent with other studies, e.g., Litjens et al. \cite{litjens2012interpatient} reporting similar variation in the peripheral zone of the prostate.  Using $b$-values up to \SI{2000}{mm^2/s} and an endorectal coil in addition to pelvic phased-array coil, Lemke et a. \cite{lemke2011diffusion} have shown higher kurtosis ($K$), as defined by the signal equation: $\ln (S/S_0) = -b D_{(2)} + b^2 D_{(2)}^{2} K,$ lower ADC and diffusion coefficient ($D_{(2)}$) values for the malignant regions in comparison to healthy ones.  Similar results for both variables have also been demonstrated in \cite{rosenkrantz2012prostate} using high $b$-values and a pelvic coil only. Sensitivity and contrast between malignant and healthy areas were higher when assessing using $K$ or $D_{(2)}$ in compared to the ADC.  These studies suggest that DWI with higher $b$-values may be clinically more advantageous, in line with PIRADS (v2), which suggests maximum $b$-values of \SI{1400}{s/mm^2} or greater.  While we see no evidence of kurtosis in our patient cohort with maximum $b$-values of \SI{1000}{mm^2/s}, kurtosis may observable for such high $b$-values.

Accurate quantification of the rectified noise floor is critical to avoid spurious kurtosis results when fitting diffusion signals, especially for high $b$-values.  A good signal-to-noise ratio provides information about the quality of an image and is vital part of the quality assurance of an MRI system \cite{firbank1999comparison} but simple SNR estimation may not be sufficient to avoid inadvertent noise floor fitting that can result in inaccurate ADC values and spurious kurtosis.  Quantification of the background noise level was in some cases compromised by image artifacts, aliasing effects or too small a FOV to place a suitable ROI, but in most cases estimation of the (rectified) noise floor was possible by selecting a suitable ROI as shown in Fig.~\ref{DelineatiatedROIs113}.  The results presented in Table~\ref{Table1} indicate that the level of background noise for different patients is relatively constant, provided the same protocol and coils are used, as expected.  The noise level for acquisitions with $b=0$ was slightly greater than for higher $b$-value images due to overall increased signal strength in the former case, but the differences were insignificant.  Similarly, there was little variation of the noise level for different slices and the overall distribution of the background noise level for all slices and $b$-values followed a normal distribution with small variance.  In our case the maximum value of the noise level observed was 7.3 and setting the threshold of significance to 10, i.e., discarding any data points with signal values below 10, appeared sufficient to avoid noise-floor fitting effects for normal and suspicious region fits as well as single voxel fits.

The single voxel analysis further enables studying the distribution of the ADC values within the whole prostatic region, producing ADC maps (Fig.~\ref{ADCmap113} (a)) and histograms that can aid clinical diagnosis by comparing the ADC values for a suspicious region with the overall distribution for a particular patient.  Although the range of ADC values for different patients is subject to variation we observe significant uniformity in the distribution of ADC for each patient, as shown in Fig.~\ref{ADChist113} (a) for example.  We also assessed the quality of the linear regression fits.  The vast majority of our data resulted in fits with $R^2$ values very close to 1 (Fig.~\ref{ADChist113} (b), for example) suggesting that the data are well explained by a monoexponential Gaussian diffusion model.  The linearity of signal decay, i.e., absence of evidence of kurtosis, is further exemplified in waterfall plots of the logarithm of the signal vs $b$-value for all voxels, as shown in Fig.\ref{ADCmap113} (b).  As the lowest ADC values for all $b$-values are significantly higher than the noise floor there should be no contamination of the signal by the noise floor even for high $b$-values.

This study is limited by small numbers and lack of histopathological correlation with the mp-MRI finding.  Not all patients went on to have a targeted biopsy, which would have allowed us to confirm whether any areas of concern were cancer or not.  We currently have funding for an extension of this study for a cohort of patients with a raised PSA and/or abnormal digital rectal examination who will have biopsies of the prostate with targeting of any abnormal areas based on mp-MRI to answer to this important question. In addition there is a limited range of $b$-values used for DWI. We would expect to find evidence of non-Gaussian diffusion at higher $b$-values and further study of the effect of the rectified noise floor on the reliability of non-linear regression fits for high $b$-values would be desirable.

\section{Conclusions}

Although further studies with a larger patient cohort are required to assess the reliability of T2w and DWI alone for early-stage prostate cancer screening examinations, our preliminary results suggest that DWI combined with T2w could be a useful clinical tool for prostate cancer assessment and management.  It could improve the efficiency of MRI-based screening protocols, reducing scan times and facilitating reporting by radiologists, especially when combined with artificial intelligence.

Our finding that DCE results correlated poorly with overall PIRADS scores and demonstrated no added value in this study may be due to DCE characteristics being mostly of value in the assessment and staging of more advanced and larger tumors. Though not the subject of this paper, the same likely applies to MRS.  The limited spatial resolution and partial volume effects combined with overall noisy spectra make it difficult to detect spectroscopic biomarkers for prostate cancer for early-stage, small volume lesions.  MRS is likely to be more useful in the assessment of larger tumors.

\section{Acknowledgements}
We thank the Swansea Prostate Cancer Research Fund and the College of Science and the NHS for funding this work.

\bibliographystyle{elsarticle-num}
\bibliography{Bibliography}

\end{document}